\documentclass[amssymb,preprint,preprintnumbers,nofootinbib,superscriptaddress]{revtex4}
\usepackage{amsmath}
\usepackage{graphicx}
\graphicspath{{Figures/}}
\usepackage{latexsym}
\usepackage{amsfonts}
\usepackage{url,hyperref}
\usepackage{bm}
\usepackage{textcomp}
\usepackage{color}
\usepackage{bbm}
\usepackage{slashed}
\usepackage{caption}
\usepackage{epstopdf}
\usepackage{subcaption}
\usepackage{placeins}
\usepackage{color}
\usepackage{cancel} 
\usepackage[normalem]{ulem}
\usepackage{tcolorbox}
\usepackage{soul}
\begin{document}

\title{The gravitational wave echoes from the black hole with three-form fields}

\author{Natthason Autthisin}
\email{natthasorn\_ut@kkumail.com}
\affiliation{Khon Kaen Particle Physics and Cosmology Theory Group (KKPaCT), Department of Physics, Faculty of Science, Khon Kaen University, Khon Kaen, 40002, Thailand}

\author{Supakchai Ponglertsakul}
\email{supakchai.p@gmail.com}
\affiliation{Strong Gravity Group, Department of Physics, Faculty of Science, Silpakorn University, Nakhon Pathom 73000, Thailand}

\author{Daris Samart}
\email{darisa@kku.ac.th}
\affiliation{Khon Kaen Particle Physics and Cosmology Theory Group (KKPaCT), Department of Physics, Faculty of Science, Khon Kaen University, Khon Kaen, 40002, Thailand}


\begin{abstract}

\textcolor{black}{
In this work, we study of massless three-form black hole, where the three-form fields are higher $p$-form gauge fields with $p=3$. These give rise to the Schwarzschild-de Sitter (Sch-dS)-like solution through an effective cosmological constant represented by $a_1$. We analyze this solution under gravitational perturbations and find that it exhibits a single-peak potential. For this case, no echoes are produced. Furthermore, we consider the massive case of the three-form fields by introducing a Stueckelberg field to restore gauge invariance and to investigate its effect on GWs at late times. In this case, the potential exhibits a double-peak structure, with the modified potential appearing beside the gravitational perturbation potential. We also examine the impact of the relevant parameters as well as the influence of the parameter $c_0$, which arises from the equation of motion of the Stueckelberg field. For a large value of $a_1$, the two peaks of the potentials are close together, while $c_0$ affects the amplitude and decay rate of the time-domain waveform, resulting in no echoes. For small values of $a_1$, the peaks of the potentials are widely separated and $c_0$ influences both the phase and the amplitude of the echoes. In addition, the quasinormal frequencies of the black hole are also calculated using both the WKB and Prony methods. As results, these provide a potential avenue for testing deviations from GR and probing possible signatures of quantum gravity through future GWs observations.}

\end{abstract}

\maketitle

\newpage

\section{Introduction}

The modification of gravitational theory and the empirical testing of the predictions derived from GR constitute essential areas of research in the fields of theoretical physics and astrophysics.  
One of the most fascinating predictions of GR is the existence of GWs. In 2016, LIGO successfully detects gravitational waves from a binary black hole merger \cite{abbott2016observation}.  It has opened a new era in black hole physics and the testing of theories of gravity in the strong-field regime. Recently, NANOGrav detected the existence of a stochastic gravitational wave background from pulsar timing array (PTA) observations in the nanohertz band \cite{arzoumanian2018nanograv}. This discovery has a significant impact on the field of gravitational wave astronomy and significantly improves observational techniques. 

In general, gravitational waves can be regarded as a perturbation of the gravitational field. The study of perturbation around black holes is firstly pioneered by Regge and Wheeler in 1957 \cite{regge1957stability} where they consider gravitational perturbation around Schwarzschild black hole. Their result is later extended to electrically charged Reissner-Nordstr\"om black hole \cite{PhysRevD.9.860} and Kerr black hole \cite{PhysRevLett.29.1114,Teukolsky:1973ha} for gravitational and electromagnetic perturbations. Another prominent result from the study of black hole perturbation is discovered by Vishveshwara in 1970 \cite{Vishveshwara:1970zz}. He studies evolution of gravitational wave package around fixed Schwarzschild spacetime. It is found that the late time signal is domiated by damping oscillation frequency. Moreover, the oscillation and damping frequencies are solely characterized by one black hole parameter i.e., black hole's mass in case of the Schwarzschild black hole (Sch-BH). These oscillations are known as QNMs and corresponding frequencies are quasinormal frequencies. It is expected that a late time signal of GWs from a binary black hole merger must be dominated by damped oscillation or a ringdown phase. At late time, a gravitational perturbation can be treated at a linearized level and therefore the oscillation modes will be described by the QNMs. For this reason, the study of QNMs has attracted significant attentions and become a very crucial tool to explore physical properties of black holes \cite{Konoplya:2011qq}, hairy black holes \cite{Dolan:2015dha,Ponglertsakul:2016fxj,Promsiri:2023yda} and compact objects \cite{Kokkotas:1999bd,Volkel:2018hwb,Aneesh:2018hlp,DuttaRoy:2019hij,Churilova:2019qph,Ponglertsakul:2022vni}. Additionally, several works have been devoted to extracting QNMs signal from the gravitational waves \cite{nollert1999quasinormal,berti2007mining,cardoso2016gravitational,cardoso1709tests}.
In 2000, Ferrari and Kokkotas \cite{ferrari2000scattering} extend the calculation of the time-domain perturbation equation of relativistic stars and find the emergence of series of late-time signals after the ringdown phase. This series of late-time signals is called gravitational wave echoes. Gravitational wave echoes are expected to be detected in the future \cite{testa2018analytical,liu2021echoes}. Theoretically, the existence of gravitational wave echoes arises from the presence of double-peaked effective potentials. The bouncing of perturbations between the double-peaked effective potentials causes the echoes to occur. Various ECOs models, particularly black holes and wormholes, have been shown to manifest double-peaked potentials and generate echoes in Refs.\cite{wang2020echoes,hui2024echoes,sang2023echoes,dong2021gravitational,guo2022echoes,liu2021echoes,dutta2020revisiting,ou2022echoes,qian2024late}. For perturbations, some ECOs models naturally yield a potential with two peaks. In other cases, such as massive gravity black holes in massive gravity \cite{dong2021gravitational}, the Stueckelberg field is employed to modify the potential to create a double-peaked structure. 

At present, there is ongoing development and a search for black hole models that are consistent with observational results. In cosmology, many cosmological problems can be described by scalar fields. In the context of ECOs, there are many models represented by scalar fields. In string theory, there exists three-form fields that have the same degrees of freedom as a scalar field \cite{groh2013duality,gubser2000supersymmetry}. Three-form fields can be used to describe inflation, structure formation, and may also serve as a candidate for dark energy in explaining the expansion of the universe \cite{koivisto2010three,koivisto2009inflation,germani2009scalar,de2012stability,kumar2014inflation,chakraborty2020dynamical}. Screening solutions involving three-form fields that are conformally coupled to matter have also been examined in \cite{barreiro2017screening}. Moreover, three-form fields have been used to study black hole and wormhole \cite{barros2018wormhole,barros2020black,tangphati2024magnetically,samart2023gravitational}. This work study gravitational perturbation of black hole supported by three-form fields. The solution for a black hole with a three-form fields have already been found in \cite{barros2020black}. It was shown that the solution has the Sch-dS-like form. It includes the existence of an effective cosmological constant, which is advantageous for exploring the expanding universe. 

Typically, perturbations in de Sitter spacetime have a single-peak potential and do not produce echoes \cite{dubinsky2024overtones,fernando2017electromagnetic}. In this work, we introduce a Stueckelberg field, inspired by the massive three-form fields case, to add an extra contribution to the effective potential and generate a double-peak structure. We use the similar manner as done in Ref.\cite{dong2021gravitational} to obtain gravitational wave echoes. Furthermore, we investigate how the amplitude of both peaks influence the echoes. 

The paper is organized as follows. In section \ref{threeform formalism}, we review general formalism of three-form fields in Einstein gravity. In section \ref{3form bh}, based on a static and spherically symmetric background, we first derive the gravitational field equations and then obtain the black hole solution for zero potential. The gravitational perturbation, time-domain wave equation, method for calculating QNMs and modification of the effective potential will be discussed in section \ref{gravpert}. Then, in section \ref{results}, we show numerical calculations of both single and double-peak potentials, time-domain strain, and QNMs of the three-form black hole for both massless and massive cases. Finally, we summarize our key results in section \ref{conclusion}.

\section{the three-form filed : general formalism} \label{threeform formalism}
In this section, we provide a brief review of the three-form fields. The action for Einstein gravity coupled to a standard three-form fields are given by \cite{barros2020black}
\begin{eqnarray}
S = \int \,d^4x \sqrt{-g} \left[ \frac{1}{2\,\kappa^2}\,R -\frac{1}{48}\,F_{\mu\nu\rho\sigma}\,F^{\mu\nu\rho\sigma} - V(A^2) \right],
\label{action}
\end{eqnarray}
where $g=\mathrm{det}\, g_{\mu\nu}$,\, $\kappa^2 \equiv 8\pi\,G$,\, $A^2 = A_{\mu\nu\rho}\,A^{\mu\nu\rho}$ and $R$ is a Ricci scalar. Additionally, $F_{\mu\nu\rho\sigma}$ is a field strength tensor of the three-form fields, $A_{\mu\nu\rho}$. It is defined as
\begin{eqnarray}
    F_{\mu\nu\rho\sigma} = \nabla_\mu\,A_{\nu\rho\sigma} - \nabla_\sigma\,A_{\mu\nu\rho} + \nabla_\rho\,A_{\sigma\mu\nu} - \nabla_\nu\,A_{\rho\sigma\mu}.
\label{3-form-stength}
\end{eqnarray}
In addition, $V(A)$ is the potential of the three-form fields. Varying the action in (\ref{action}) with respect to $g^{\mu\nu}$, the Einstein field equation (EFE) is given by
\begin{eqnarray}
G_{\mu\nu} = \kappa^2\,T_{\mu\nu}\,,
\label{efe}
\end{eqnarray}
where $G_{\mu\nu} \equiv R_{\mu\nu} -\frac12\,g_{\mu\nu}\,R$\,. The energy-momentum tensor of the three-form fields reads
\begin{eqnarray}
T_{\mu\nu} \equiv \frac{1}{6}\,F_{\mu}^{~\,\rho\sigma\tau}\,F_{\nu\rho\sigma\tau} + 6\frac{\partial\,V(A^2)}{\partial (A^2)}\,A_{\mu}^{~\,\rho\sigma}\,A_{\nu\rho\sigma} + g_{\mu\nu}\left[\frac{1}{48}\,F_{\alpha\beta\rho\sigma}\,F^{\alpha\beta\rho\sigma} - V(A^2) \right]\,.
\label{EMT-3-form}
\end{eqnarray}
The equation of motion of the three-form is obtained by varying with respect to the $A_{\mu\nu\rho}$. One finds 
\begin{eqnarray}
\nabla_\sigma F^{\sigma\mu\nu\rho} - 12\frac{\partial\,V(A^2)}{\partial (A^2)}\,A^{\mu\nu\rho} = 0\,. \label{eom 3form}
\end{eqnarray}
The three-form gauge field can be represented as a product of the one-form dual vector and Levi-Civita tensor 
\begin{equation}
A_{\mu\nu\rho}=\sqrt{-g}\epsilon_{\mu\nu\rho\sigma}B^{\sigma}, \label{3-form-dual}   
\end{equation}
where the $B^\sigma$ is the one-form dual vector of the three-form fields, $A_{\mu\nu\rho}$.
The one-form dual vector can be written using the radial ansatz  \cite{barros2020black}
\begin{equation}
B^{\sigma}=\left(0,\zeta(r),0,0\right)^{\intercal}\,,\label{3-form-dual-comp}  
\end{equation}
where $\zeta(r)$ is a generic parametrization of the three-form fields. The three-form field function $\zeta$ will be determined from the EFE \eqref{efe} and equation of motion \eqref{eom 3form}.

With the ansatz \eqref{3-form-dual-comp}, the kinetic term of the three-form field may be expressed explicitly as  
\begin{eqnarray}
    -\frac{1}{48}F^2=-\frac{1}{48}F_{0123}F^{0123}=\frac{1}{2}\left( \nabla_{\mu} B^{\mu}  \right)^2 \label{kinetic 3form}.
\end{eqnarray}
In the following section, we shall consider static spherically symmetric solution of Einstein-three-form field theory.

\section{Black hole with three-form fields}\label{3form bh}
In this section, we calculate the EFE from the three-form field and obtain the black hole solution supported by the three-form field.
\subsection{Black hole metric and field equations}
In this work, we examine the following static and spherically symmetric ansatz 
\begin{eqnarray}
    ds^2=-e^{\alpha(r)} dt^2+e^{\beta(r)} dr^2+r^2 d\Omega^2, \label{static spherical metric}
\end{eqnarray}
where $d\Omega^2=d\theta^2+\sin^2\theta d\phi^2$.
On this background geometry, the invariant $A^2$ is given by
\begin{eqnarray}
    A^2=-6 e^{\beta(r)}\zeta(r)^2,
\end{eqnarray}
and the term expressing the kinetic energy of the three-form is provided explicitly by 
\begin{eqnarray}
F^2=-6\bigg[\zeta\left(\alpha'+\beta'+\frac{4}{r} \right)+2\zeta' \bigg]^2, \label{F^2}
\end{eqnarray}
where a prime denotes the derivative with respect to the radial coordinate. EFE (\ref{eom 3form}) can now be written in terms of $\zeta$, as
\begin{eqnarray}
    2\zeta''+\left( \alpha'+\beta'+\frac{4}{r} \right)\zeta'+\left( \alpha''+\beta''-\frac{4}{r^2} \right)\zeta+2\frac{\partial V}{\partial \zeta} =0\,.  \label{eom 3form new}
\end{eqnarray}
From 
the energy-momentum tensor (\ref{EMT-3-form}) and the metric ansatz (\ref{static spherical metric}), we obtain the components of energy-momentum tensor of the three-form fields as 
\begin{align}
    T^{t}_{t}&=-\rho=\frac{F^2}{48}-V+\frac{\partial V}{\partial \zeta}, \\
    T^{r}_{r}&=p_r=\frac{F^2}{48}-V, \\
T^{\theta}_{\theta}&=T^{\phi}_{\phi}=p_t=T^{t}_{t}. \label{EMT components}
\end{align}
We identify $\rho$, $p_r$ and $p_t$ with the energy density, the radial pressure, and the tangential pressure, respectively, in term of the three-form. We notice that $\rho=-p_t$. 
Now by setting $\kappa=1$ for simplicity, three components of EFE \eqref{efe} read 
\begin{eqnarray}
   \label{efe tt} 
   \frac{e^{-\beta}}{r^2}\left(1-r\beta'-e^{\beta}\right)&=&\frac{F^2}{48}-V+\frac{\partial V}{\partial \zeta},\\ \label{efe rr}
    \frac{e^{-\beta}}{r^2}\left(1+r\alpha'-e^{\beta}\right)&=&\frac{F^2}{48}-V,\\
    \label{efetheta}
    \frac{e^{-\beta}}{2}\bigg[ \alpha''+\left(\frac{1}{r}+\frac{\alpha'}{2}  \right)\left(\alpha'-\beta' \right) \bigg]&=&\frac{F^2}{48}-V+\frac{\partial V}{\partial \zeta}.
\end{eqnarray}
Combining \eqref{efe tt} and \eqref{efe rr}, we find 
\begin{eqnarray}
    \alpha'+\beta'= -re^{\beta} \zeta\frac{\partial V}{\partial \zeta} .\label{alpha beta diff}
\end{eqnarray}
By considering \eqref{efe tt} and \eqref{efetheta}, we obtain
\begin{eqnarray}
    \frac{2}{r^2}\left( 1-r\beta'-e^{\beta} \right)=\alpha''+\left( \frac{1}{r}+\frac{\alpha'}{2}   \right) \left( \alpha'-\beta'  \right)\,. \label{alpha2diff}
\end{eqnarray}
With a given potential $V$, there are three coupled differential equations \eqref{eom 3form new},\eqref{alpha beta diff}--\eqref{alpha2diff} for three undetermined functions $\alpha(r),\,\beta(r)$ and $\zeta(r)$. In the next subsection, we will solve this equation system for the black hole solution.

\subsection{Black hole solution with three-form fields}
\begin{figure}[h]
    \includegraphics[scale=0.5]{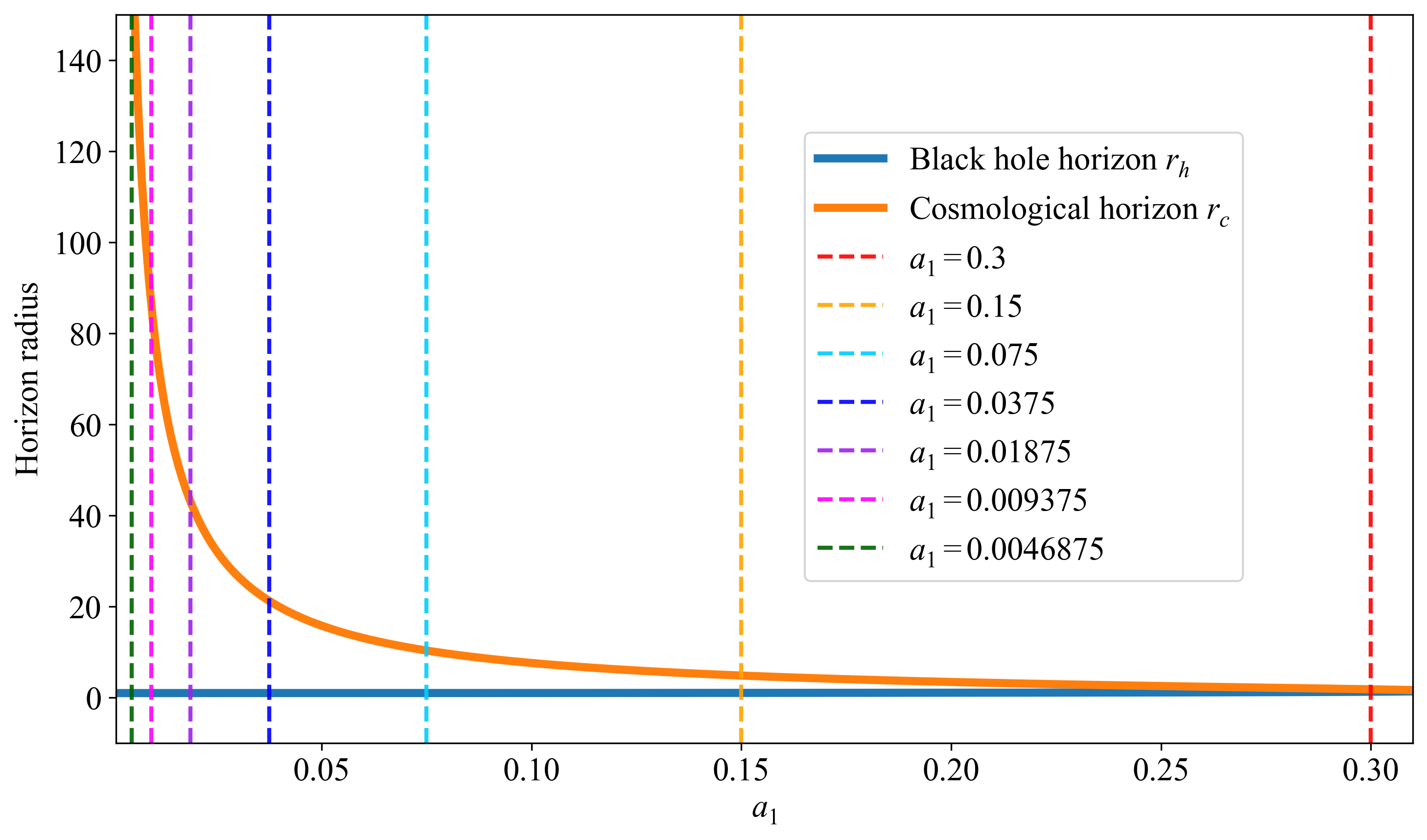}
\caption{Horizon radius as functions of the parameter $a_1$, with $M=1/2$. The blue and orange curves represent the black hole horizon radius $r_h$ and the cosmological horizon radius $r_c$, respectively. Vertical dashed lines indicate specific values of $a_1$ used in the analysis throughout this work.}
 \label{rh and rc}
\end{figure}

To obtain a static spherically symmetric black hole solution, we shall set the three-form field's potential $V=0$. This is effectively equivalent to the massless three-form fields. Thus from \eqref{alpha beta diff}, we get
\begin{eqnarray}
    \alpha=-\beta, \label{alpha beta 0}
\end{eqnarray}
This greatly simplified our set of equations. Hence, \eqref{alpha2diff} becomes
\begin{eqnarray}
    \alpha''+\alpha'^2=\frac{2}{r^2}\left( 1-e^{-\alpha(r)}  \right)\,.\label{main alpha beta}
\end{eqnarray}
This equation can be solved by setting $\alpha=\ln f$. 
Thus, we obtain the solution
\begin{eqnarray}
    f(r)=1+\frac{c_1}{r}+c_2 r^2, \label{metricfunction}
\end{eqnarray}
where $c_1$ and $c_2$ are the constants of integration. The parameter $c_1$ can be determined by comparing it with the Schwarzschild solution, i.e., $c_1=-2M$, where $M$ is the mass of the black hole. To determine $c_2$, we consider the following. By using, \eqref{alpha beta 0}, the equation of motion of the three-form fields \eqref{eom 3form new} can be re-casted into 
\begin{eqnarray}
    \zeta''+\frac{2}{r}\zeta'-\frac{2}{r^2}\zeta=0\,,\label{zeta eom}
\end{eqnarray}
which yields the following solution 
\begin{eqnarray}
    \zeta(r)=a_1 r + \frac{a_2}{r^2} \label{zeta}
\end{eqnarray}
where $a_1$ and $a_2$ are arbitrary constants of integration. Therefore, the kinetic term of the three-form field can be written explicitly in terms of constant $a_1$
\begin{eqnarray}
    F^2=-216 a^2_{1}\,.
\end{eqnarray}
the constant $c_2$ can be written in terms of $a_1$ by putting the exact form of $f(r)$ into \eqref{efetheta}. As a result, we obtain $c_2=-3a_1^2/2$. Therefore, we can express $f(r)$ in an Sch-dS-like solution. It reads
\begin{eqnarray}
    e^{\alpha}=e^{-\beta}=f(r)=1-\frac{2M}{r}-\frac{3}{2}a_{1}^2 r^2,
\end{eqnarray}
where $a_{1}$ can be related to an effective cosmological constant ($\Lambda_{\rm eff}$) emerging from the three-form field via $\Lambda_{\rm eff} \equiv 9a_1^2/2$. We can rewrite the matric (\ref{static spherical metric}) as
\begin{eqnarray}
    ds^2=-\left(1-\frac{2M}{r}-\frac{3}{2}a_{1}^2 r^2\right)dt^2+\left(1-\frac{2M}{r}-\frac{3}{2}a_{1}^2 r^2  \right)^{-1}dr^2+r^2 d 
    \Omega^2\label{bh metric full}\,.
\end{eqnarray}
The Sch-dS spacetime has two horizons: the black hole horizon is at $r = r_h$, and the cosmological horizon is at $r = r_c$, where $r_c > r_h$. The function $f$ vanishes at $r_h, r_c,$ and $r_0 = -(r_h + r_c)$. The function $f$ can be expressed as 
\begin{eqnarray}
    f(r)=\frac{3a_1^{2}}{2r}(r-r_h)(r_c-r)(r-r_0).
\end{eqnarray}
The root of cubic equations can be expressed in analytic form as follows
\begin{eqnarray}
    \label{rh}r_h&=&\frac{2\sqrt{2}}{3a_{1}}\cos\left( \frac{1}{3}\arccos\left( -\frac{9M a_{1}}{\sqrt{2}}  \right)+4\pi   \right), \\
    \label{rc} r_c&=&\frac{2\sqrt{2}}{3a_{1}}\cos\left( \frac{1}{3}\arccos\left( -\frac{9M a_{1}}{\sqrt{2}}  \right) \right),\\
    \label{r0} r_0&=&\frac{2\sqrt{2}}{3a_{1}}\cos\left( \frac{1}{3}\arccos\left( -\frac{9M a_{1}}{\sqrt{2}}  \right)+2\pi   \right)\,.
\end{eqnarray}
In this work, we consider the value of $a_1$ from the near-extremal case to the case where the black hole horizon $r_h$ and the cosmological horizon $r_c$ are widely separated, as shown in Fig. \ref{rh and rc}. Specifically, the value of $a_1$ starts from $0.3$ and is successively halved down to $0.0046875$, as shown by the red to green lines in Fig. \ref{rh and rc}.

We also introduce the surface gravity $k_i$ associated with the horizon $r = r_i$, as defined by the relation $k_i =
\frac{1}{2}|df /dr|_{r=r_i}$. Then, we have 
\begin{eqnarray}
   k_h&=& \frac{3a_{1}^2(r_c-r_h)(r_h-r_0)}{4 r_h},\\
    k_c&=&\frac{3a_{1}^2(r_0-r_c)(r_c-r_h)}{4 r_c},\\
    k_0&=&\frac{3a_{1}^2(r_c-r_0)(r_0-r_h)}{4 r_0}\label{surface gravity}\,.
\end{eqnarray}
We define the tortoise coordinate $r_{\ast}=\int f^{-1} dr$. From the surface gravity $k_i$, we can express $f^{-1}$ as 
\begin{eqnarray}
    f^{-1}=\frac{1}{2k_h(r-r_h)}+\frac{1}{2k_c(r_c-r)}+\frac{1}{2k_0(r-r_0)}.
\end{eqnarray}
Therefore, the tortoise coordinate $r_*$ in Sch-dS spacetime can be obtained exactly \cite{brady1999radiative}
\begin{eqnarray}
r_*=\frac{1}{2k_h}\mathrm{ln}\left( \frac{r}{r_h}-1\right)-\frac{1}{2k_c}\mathrm{ln}\left( 1-\frac{r}{r_c} \right)+\frac{1}{2k_0}\mathrm{ln}\left( 1-\frac{r}{r_0} \right)\,.
\end{eqnarray}
In the next section, we will explore gravitational perturbation in black hole spacetime.

\section{gravitational perturbation}\label{gravpert}
In the previous section, we obtain the three-form black hole solution. In this section, we study perturbations in the black hole spacetime. The type of perturbation examined in this work is a gravitational perturbation. We analyze the behavior of effective potentials and introduce an additional term into the potential. This additional term will play the role as a mass and an interaction term for the three-form fields. We also examine the time-domain profiles of gravitational perturbation around three-form black holes. In addition, we also determine the QNMs of the three-form black hole. 

We start by considering the following linear perturbation on the spacetime metric 
\begin{eqnarray}
    g_{\mu\nu}=g^0_{\mu\nu}+h_{\mu\nu}\,.
\end{eqnarray}
where $g^0_{\mu\nu}$ is background metric \eqref{bh metric full}. The small perturbation is denoted by rank-2 tensor $h_{\mu\nu}$. From \cite{regge1957stability,otsuki1991gravitational}, the canonical form of $h_{\mu\nu}$ for the gravitational perturbations in the Regge-Wheeler gauge is given by
\begin{eqnarray}
   h_{\mu\nu} = \begin{bmatrix}
0 & 0 & 0 & h_0(t,r) \\
0 & 0 & 0 & h_1(t,r)\\
0 & 0 & 0 & 0\\
h_0(t,r) & h_1(t,r) & 0 & 0
\end{bmatrix} \sin\theta\left({\frac{\partial}{\partial \theta}}\right) P_l(\cos\theta)\,,
\label{RW-gauge}
\end{eqnarray}
where $P_l(\cos\theta)$ is the Legendre polynomial of order $l$ with two unknown functions $h_0(t,r)$ and $h_1(t,r)$. 


\subsection{Time-domain : wave-like equation}
To obtain the Schr\"odinger-like equation for the gravitation perturbation, we define time-domain wavefunction as expressed below
\begin{eqnarray}
    \Psi(t,r)=f(r)\frac{h_1(t,r)}{r}\label{ansatz gw},
\end{eqnarray}
Having use the standard linearization of EFE with Eqs.(\ref{RW-gauge}) and (\ref{ansatz gw}), one finds the Schr\"odinger-like equation as 
\begin{eqnarray}
    \left(\frac{\partial^2 }{\partial t^2}-\frac{\partial^2 }{\partial r_*^2}+V_{\mathrm{GP}}(r)\right)\Psi(t,r_*)=0\,,\label{wavelike eq}
\end{eqnarray}
where we define the effective potential from gravitational perturbation $V_{\mathrm{GP}}$ as
\begin{eqnarray}
    V_{\mathrm{GP}}(r) = f(r)\left(\frac{l(l+1)}{r^2}-r\frac{d}{dr}\left(\frac{f(r)}{r^2}\right)-\frac{2}{r^2}   \right).
\end{eqnarray}
We note that the $h_0(t,r)$ can be eliminated via the linearized EFE. 

\subsubsection{The modified potential from massive three-form fields} \label{stuekelberg}
In the previous section, black hole solution with massless three-form fields is obtained. However, one can investigate how massive three-form fields affects to gravitational waves. In this section, we examine the effect of massive three-form fields to gravitational waves. We are interested in how the energy-momentum tensor of mass term modifies the axial perturbation potential to produce a double-peak structure of the effective potential.
From previous work in \cite{dong2021gravitational} and standard gravitational perturbation analysis, the gravitational perturbation equation receives additional potential terms if the background matter fields couple to the metric in a non-trivial radial profile.

We start by considering the action of a massive three-form, which can be written as
\begin{eqnarray}
    S = \int d^4x \sqrt{-g} \left( \frac{1}{2\kappa^2} R - \frac{1}{48} F_{\mu\nu\rho\sigma}F^{\mu\nu\rho\sigma} - \frac{1}{2} m_A^2 A_{\mu\nu\rho}A^{\mu\nu\rho} \right),
\end{eqnarray}
where $m_A << 1$ is the small mass of three-form fields. According to the one-form dual vector of the three-form in Eq.(\ref{3-form-dual}) and Eq.(\ref{kinetic 3form}), the action become massive spin-1 field as 
\begin{eqnarray}
    S = \int d^4x \sqrt{-g} \left( \frac{1}{2\kappa^2} R + \frac{1}{2} (\nabla_\mu B^\mu)^2 - \frac{1}{2} m_A^2 B^\mu B_\mu \right).
\end{eqnarray}
In contrast to the scalar perturbation where scalar field's mass can be introduced directly into the Lagrangian, adding mass term to vector field will break gauge invariance. In addition, the mass term leads to pathological problems such as ghost and unstable longitudinal modes \cite{clough2022problem,dolan2018instability,east2017superradiant}.

Similarly, the first massive graviton theory introduced by Fierz and Pauli \cite{Fierz:1939ix} is known to suffer from the van Dam-VeltmanZakharov (vDVZ) discontinuity \cite{vanDam:1970vg,Zakharov:1970cc} where GR is not recovered in the massless graviton limit and the Boulware-Deser ghost \cite{Boulware:1972yco} which leads to additional unphysical degree of freedom to the theory. The pathological-free theory of massive gravity is proposed by de Rham, Gabadadze, and Tolley (dRGT) \cite{deRham:2010ik,deRham:2010kj}. In dRGT massive gravity, graviton mass term is described in term of non-linear Stueckelberg fields. The Stueckelberg fields are not fundamental but rather auxiliary fields to help preserve diffeomorphism invariance in massive gravity theory. By treating massive gravity as a small correction to GR, diffeormophism invariance remains intact at the first order perturbation. Remarkebly, the correction term can be introduced as coupling between the perturbation and background Stueckelberg field \cite{Dong:2015qpa}. This effectively renders a mass term for the graviton. In the same manner as the massive graviton, to restore $U(1)$ gauge invariance in the case of dual vector of massive three-form fields, we introduce a scalar Stückelberg field $\pi$ via
\begin{eqnarray}
    B^\mu \rightarrow B^\mu+ \partial^\mu \pi.
\end{eqnarray}
The action reads
\begin{eqnarray}
    S = \int d^4x \sqrt{-g} \left( \frac{1}{2\kappa^2} R + \frac{1}{2} \left(\nabla_\mu( {B}^\mu+\partial^\mu \pi)\right)^2 - \frac{1}{2} m_A^2 \left(B^\mu+\partial^\mu \pi\right)^2 \right),
\end{eqnarray}
where the action is now invariant under the gauge transformation
\begin{equation}
B^\mu \rightarrow B^\mu+\partial^\mu \chi, \quad \pi \rightarrow \pi-\chi,
\end{equation}
where the scalar function, $\chi=\chi(x^\mu)$ is an arbitrary function.  
The action shows the Stueckelberg scalar $\pi$ couples directly to the vector field via the gauge invariant mass term $B^\mu\partial_\mu \pi$ and $(\partial \pi)^2$. This is exactly analogous to the coupling between the scalar Stueckelberg field $\phi^a$ and the
metric tensor in massive gravity. 

To determine the solution of the Stueckelberg field, $\pi$, one can evaluate from the the EOM of $\pi$, and it reads,
\begin{eqnarray}
    \nabla_\mu\left(B^\mu+\partial^\mu \pi\right)&=&0,
\end{eqnarray}
where the components of term $(B^\mu+\partial^\mu \pi)$ can be expressed as 
\begin{eqnarray}
    B^\mu+\partial^\mu \pi=\left(\partial^t \pi, \zeta(r)+\partial^r \pi, 0,0\right).
\end{eqnarray}
Considering the $\pi$ field in the static limit, we obtain
\begin{eqnarray}
    \nabla_\mu\left(B^\mu+\partial^\mu \pi\right)
   &=& \frac{1}{r^2} \partial_r\left(r^2 f(r)\big[\zeta(r)+\partial_r \pi\big]\right). \nonumber 
\end{eqnarray}
From the EOM of $\pi$, one finds
\begin{eqnarray}
    \partial_r\left(r^2 f(r)\left[\zeta(r)+\partial_r \pi(r)\right]\right)=0.
\end{eqnarray}
The derivative of $\pi(r)$ with respect to $r$, $\partial_r \pi(r)\equiv \pi'(r)$ can be expressed as
\begin{eqnarray}
    \pi'(r)=\frac{c_0}{r^2 f(r)}-\zeta(r),
\end{eqnarray}
where $c_0$ is the integration constant. The component $\partial_r \pi(r)$ is important for modifying the axial perturbation potential. The mass term contributes to the energy-momentum tensor, which is given by
\begin{eqnarray}
    T_{\mu \nu}^{ (m)}=m_A^2\left(B_\mu+\partial_\mu \pi\right)\left(B_\nu+\partial_\nu \pi\right)-\frac{1}{2} m_A^2 g_{\mu \nu}\left(B^\alpha+\partial^\alpha \pi\right)^2.
\end{eqnarray}
This modifies the EFE, and therefore, backreacts on the potential governing metric perturbations.
The modified effective potential $\Delta V$ for tensor perturbations can be defined by the following expression
\begin{eqnarray}
    \Delta V(r)=\delta T_{\mu \nu}^{(\text {m })}.
\end{eqnarray}
According to Stueckelberg's mechanism, we defined a new field variable $\mathcal{A}^{\mu}$ as
\begin{eqnarray}
    \mathcal{A}^\mu \equiv B^\mu+\partial^\mu \pi,
\end{eqnarray}
then, $T^{(m)}_{\mu\nu}$ can be rewritten as
\begin{eqnarray}
    T^{(m)}_{\mu\nu} = m_A^2 \mathcal{A}_{\mu} \mathcal{A}_{\nu} - \frac{1}{2} m_A^2 g_{\mu\nu}\mathcal{A_{\alpha}}\mathcal{A}^{\alpha}
\end{eqnarray}
The modifed potential that projected onto gravitational perturbation is given by.
\begin{eqnarray}
    \Delta V \thicksim m_A^2 \mathcal{A_\mu A^{\mu}}.
\end{eqnarray}
Noting that inclusion of the mass term of the three-form fields as the modified potential of the perturbation, we still cannot obtain the double peak structure of the effective potential. In order to produce the second peak as well as to keep the modified potential invariance under the gauge transformation with the Stueckelberg mechanism, it is necessary to introduce additional terms. Accordingly, we propose the following form for the perturbed potential inspired by the Ratra–Peebles potential (inverse power law field) \cite{Peebles:1987ek,Ratra:1987rm}. It reads, 
\begin{eqnarray}
\Delta V(r) &=& m_A^2\mathcal{A}_\mu \mathcal{A}^\mu + \lambda \left( \mathcal{A}_\mu \mathcal{A}^\mu\right)^{-2} 
\nonumber\\ 
&=& m_A^2 \frac{ c_0^2}{r^4 f(r)^3} + \lambda\left(\frac{ c_0^2}{r^4 f(r)^3}\right)^{-2},
\end{eqnarray}
where $\mathcal{A}_\mu \mathcal{A}^\mu = \frac{1}{f}\left(\zeta+\pi'\right)^2$ and $\lambda$ is a coupling parameter that quantifies the strength of the new interaction term. The term $\lambda \left( \mathcal{A}_\mu \mathcal{A}^\mu\right)^{-2} $ is crucial to shaping the potential to exhibit a double peak while preserving gauge invariance. 

\textcolor{black}{We are particularly aware of the behavior of the effective potential, which asymptotically approaches zero near both the black hole and the cosmological horizons, but diverges at the horizons themselves. To maintain consistency with the condition of asymptotic flatness, we restrict our analysis to regions sufficiently close to, but not exactly at, the black hole and cosmological horizons.}

\subsubsection{Calculation of time-domain wavefunction}
We use the light cone coordinate, where $u=t-r_*$ is the retarded time and $v=t+r_*$ is the advanced time, In terms of the light cone coordinate $u$ and $v$, the wave equation \eqref{wavelike eq} becomes \cite{gundlach1994late}
\begin{eqnarray}
 4\frac{\partial^2 \Psi}{\partial u \partial v}+V_{\mathrm{eff}}(u,v)\Psi=0\,, \label{wavefunction timedomain}
\end{eqnarray}
where $V_{\mathrm{eff}}=V_{\mathrm{GP}}$ in the case of massless three-form fields, and $V_{\mathrm{eff}}=V_{\mathrm{GP}}+\Delta V$ for the inclusion of the modified potential case. To evaluate the wavefunction $\Psi(t,r_*)$, we use the following discretization scheme  \cite{gundlach1994late}. The coordinates $u$ and $v$ increases by discrete step $\Delta$. In the discrete domain, the differential equation is defined by
\begin{eqnarray}
\Psi(N)=\Psi(W)+\Psi(E)-\Psi(S)-\frac{\Delta^2}{8}V_{\mathrm{eff}}(S)\bigg[\Psi(W)+\Psi(E)\bigg]\, , \label{gunlach}
\end{eqnarray}
\textcolor{black}{where the points $N$, $S$, $E$, and $W$ refer to points on a compass on a null rectangle.
These points are defined as $N : (u+h,v+h), W : (u+h,v), E : (u,v+h)$ and $S : (u,v)$.}

\subsection{Frequency-domain : QNMs of back hole}

A linear perturbation on black hole results in an oscillation which can be described by QNMs \cite{chandrasekhar1975quasi}. A characteristic oscillation signal with corresponding complex frequencies which depend on black hole parameters. 

Black hole perturbation equation is often put into the Schr\"odinger-like equation with  effective potential \ref{wavelike eq}. This equation can be written in the Regge-Wheeler equation which solely radial-dependence. By using, the ansazt 
\begin{eqnarray}
    \Psi(t,r_*)=e^{-i \omega t}\psi(r_*)\,,\label{psi(t,r)}
\end{eqnarray}
Eq. \eqref{wavelike eq} becomes
\begin{eqnarray}
    \frac{d^2 \psi}{dr_*^2}+ \bigg[\omega^2-V_{\mathrm{\mathrm{eff}}}(r) \bigg]\psi=0, \label{regge}
\end{eqnarray}
where $\omega$ is denoted as the quasinormal frequency \cite{birmingham2001choptuik,kao2008quasinormal,dal2023quasinormal,cao2023quasinormal,he2009quasinormal,sadeghi2011quasi,musiri2003quasinormal}. Boundary condition associated with quasinormal modes of black hole are
\begin{eqnarray}
    \text{Ingoing} \quad :& \quad &\psi(r_*) \sim e^{-i \omega r_*}, r_*\rightarrow -\infty, \\
    \text{Outgoing} \quad :& \quad &\psi(r_*) \sim e^{i \omega r_*}, r_*\rightarrow \infty.
\end{eqnarray}
At event horizon, only ingoing wave is allowed while only outgoing wave is permitted at asymptotic infinity. This particular boundary condition yields discrete complex frequency $\omega=\omega_R\pm i\omega_I$. The real part describes an oscillation of a perturbed field while imaginary part determining whether it exhibits decaying or growing behavior.

There are several methods to determine quasinormal frequencies, for instance, Asymptotic Iteration Method (AIM) \cite{cho2010black,ciftci2003asymptotic}, the Continued Fraction Method (CFM) \cite{leaver1985analytic}, Chebyshev pseudospectral method \cite{boyd2001chebyshev,aragon2021massive}, etc. While the mentioned methods rely on sophisticated numerical scheme, one of the most commonly used is the Wentzel–Kramers–Brillouin (WKB) method. It is regarded as a semi-analytical method and relatively simpler comparing with the other approaches. A huge number of QNMs studies utilise the advantage of WKB method \cite{schutz1985black,PhysRevD.35.3621,konoplya2003quasinormal,zhidenko2003quasi,Burikham:2017gdm,Ponglertsakul:2018smo,matyjasek2019quasinormal,konoplya2019higher,santos1903quasi,Ponglertsakul:2020ufm,Ponglertsakul:2022vni,gogoi2023quasinormal,tangphati2024magnetically,gogoi2024constraints}.

The WKB method has an advantage when an exact form of effective potential is known. The quasinormal frequencies can be directly computed from the effective potential. 
\textcolor{black}{More importantly, the WKB approach is valid only for potentials with a single-peak structure, where the wave behavior near the maximum dominates the scattering process. In this work, double-peak potentials are also considered due to the presence of $\Delta V$.} Therefore, the WKB method might not be suitable here. To obtain the quasinormal frequencies, we extract frequencies profile from the time-domain profile of QNMs via the Prony method instead \cite{konoplya2011quasinormal,guo2022echoes,chowdhury2020echoes,vishvakarma2023shadows,berti2007mining}. 

\textcolor{black}{Thus, for a massless case, we will determine quasinormal frequencies of gravitational perturbation around three-form fields black hole via the 6th-order WKB and the Prony method. When three-form fields acquire mass via the Stueckelberg mechanism, the frequencies are obtained from the Prony method.}

\subsubsection{WKB method}
The 1st-order WKB method is firstly introduced by Schutz and Will \cite{schutz1985black} and later is extended with higher order correction terms \cite{PhysRevD.35.3621,konoplya2003quasinormal,matyjasek2019quasinormal}. The WKB method is based on the problem of waves scattering near the peak of the potential barrier $V(r_*)$ in quantum mechanics, where $\omega^2$ acts as energy. In this work, we consider the 6th-order WKB. In order to obtain the QNMs, we rewrite \eqref{regge} as 
\begin{eqnarray}
    \frac{d^2 \psi}{d r_*^2}+Q(r_*)\psi(r_*)=0\,.
\end{eqnarray}
where $Q\equiv \omega^2-V(r_\ast)$. The quasinormal frequencies can be found by solving the following formula \cite{konoplya2003quasinormal}
\begin{eqnarray}
    \frac{i Q_0}{\sqrt{2Q''_0}}-\bar \Lambda_2-\bar \Lambda_3-\bar \Lambda_4-\bar \Lambda_5-\bar \Lambda_6=n+\frac{1}{2}.\label{WKB Q}
\end{eqnarray}
$Q_0 =\omega^2-V_0$ and $V_0$ is the value of the potential at the maximum point. We denote second derivative with respect to tortoise coordinate i.e., $Q_0''=d^2Q/dr_\ast^2|_{r=r_{max}}$. An overtone number is denoted by integer $n$. The correction terms $\bar\Lambda_{2-6}$ up to 6th-order can be found in \cite{PhysRevD.35.3621,konoplya2003quasinormal}. When the effective potential $V$ is independent of $\omega$, the quasinormal frequencies are obtained as 
\begin{eqnarray}
    \omega =\sqrt{-i\bigg[\left(n+\frac{1}{2} \right)+\sum^6_{k=2} \bar \Lambda_k \bigg ]\sqrt{-2V_0''}+V_0} \, .
\end{eqnarray}
\textcolor{black}{We will employ this formula in the following analysis to obtain the quasinormal frequencies of the massless three-form black hole only.}

\subsubsection{Prony method}
To evaluate the QNMs during the ringdown phase, we fit the time-domain profile by treating it as a sum of damped exponentials
\begin{eqnarray}
    \Psi(t)\simeq \sum_{n=1}^{p} C_n e^{-i \omega_n t} \,.\label{damped}
\end{eqnarray}
However, the calculation is complicated. In 1795, Prony proposes a generalized approach for fitting a sum of damped exponentials, known as the Prony method \cite{baron1795essai}. We assume that the ringdown phase begins at $t_0=0$ and ends at $t=Nh$, where $N\ge 2p-1$. Here, $h$ is the fixed time interval between consecutive samples of the signal $\Psi(t)$, set by the data acquisition rate. Then, the relation (\ref{damped}) is satisfied for each point $m$ of the time-domain profile as
\begin{eqnarray}
    \mathrm{x}_m=\Psi(mh)=\sum_{n=1}^{p}C_ne^{-i\omega_n m h }\equiv\sum_{n=1}^{p}C_n z^m_n \,.
\end{eqnarray}
We use Prony method to determine $z_n$ from the data $\mathrm{x}_m$. Then, we calculate QNFs ($\omega_n$) from $z_n$. To find $z_n$, we define a polynomial $\Tilde{A}(z)$ as follows 
\begin{eqnarray}
    \Tilde{A}(z)=\prod_{n=1}^p(z-z_n)=\prod_{k=0}^p\alpha_k z^{p-k}, \quad \alpha_{0}=1, \label{A(z)}
\end{eqnarray}
and consider the summation
\begin{eqnarray}
    \sum_{k=0}^p \alpha_k \mathrm{x}_{m-k}=\sum_{k=0}^p\alpha_k\sum^p_{n=1}C_n z^{m-k}_n=\sum^p_{n=1} C_n z_n^{m-p} \sum_{k=0}^p \alpha_k z_n^{p-k}\,.
\end{eqnarray}
Then, we obtain
\begin{eqnarray}
    \sum_{k=1}^p\alpha_k \mathrm{x}_{m-k}=-\mathrm{x}_m.
\end{eqnarray}
Using $m=p,...,N$ in \eqref{Xax}, we obtain $N-P+1 \ge p$ linear equation for $\alpha_k$, which can be expressed in matrix form as
\begin{eqnarray}
    X\alpha=-\mathrm{x}, \label{Xax}
\end{eqnarray}
where
\begin{eqnarray}
    X = \begin{pmatrix} 
    \mathrm{x}_{p-1} & \mathrm{x}_{p-2} & \dots &\mathrm{x}_0 \\
    \mathrm{x}_{p} & \mathrm{x}_{p-1} & \dots &\mathrm{x}_{1}\\
    \vdots & \vdots & \ddots & \vdots \\
    \mathrm{x}_{N-1} & \mathrm{x}_{N-2} &\dots & \mathrm{x}_{N-p} 
    \end{pmatrix},
\quad
\alpha = \begin{pmatrix} 
    \alpha_1  \\
    \alpha_2 \\
    \vdots\\
    \alpha_p 
    \end{pmatrix},
    \quad
    \mathrm{x}=\begin{pmatrix} 
    \mathrm{x}_p  \\
    \mathrm{x}_{p+1} \\
    \vdots\\
    \mathrm{x}_N 
    \end{pmatrix}\,.
\end{eqnarray}
From the above equation, we can solve for the coefficient matrix $\alpha$ using the expression below
\begin{eqnarray}
    \alpha=-(X^{\dagger} X)^{-1} X^{\dagger} \mathrm{x},
\end{eqnarray}
where $X^{\dagger}$ is the Hermitian transpose of $X$. Substituting $\alpha$ into \eqref{A(z)}, we obtain the roots $z_n$ of the polynomial $\Tilde{A}(z)$. Finally, the QNMs can be calculated using the relation below
\begin{eqnarray}
    \omega_n=\frac{i}{h}\ln{(z_n)} \,, \label{prony omega}
\end{eqnarray}
with $\omega_n$ determined from Eq. (\ref{prony omega}). The QNMs are fully characterized, completing the Prony method reconstruction of the ringdown signal as a sum of damped exponentials Eq. (\ref{damped}).

\section{numerical results}\label{results}
In this section, we investigate the behavior of the effective potential ($V_{\mathrm{eff}}$), including the gravitational perturbation potential ($V_{GP}$) and the potential from the Stueckelberg field ($\Delta V$). \textcolor{black}{Without $\Delta V$, the effective potential of gravitational perturbation is single-peak type. By introducing the modified potential produced by the stueckelberg field, the effective potential becomes double-peak type. Moreover, the inclusion of the stueckelberg field also introduces the mass term of the three-form fields.} 
Exploring the behavior of the potential allows us to analyze and calculate the time-domain profile of the QNMs. In order to numerically solve Eq. \eqref{wavefunction timedomain}, we consider the Gaussian profile as the initial data
\begin{eqnarray}
    \Psi(0,v)=\mathrm{exp}\left({-\frac{(v-v_c)^2}{2\sigma^2}}\right)\quad \text{and} \quad \Psi(u,0)=0 ,\label{gaussian}
\end{eqnarray}
where the Gaussian profile is centered at $v_c=10$ and has a  width $\sigma=3$. The initial position of the Gaussian profile, $v_c$, is placed near the peak of $V_{\mathrm{GP}}.$ Here, we particulary focus on the fundamental mode $(n=0)$ and spherical harmonics with $l=2$ due to their crucial contribution to the ringdown gravitational waves following the merger of binary black holes \cite{giesler2019black,sago2021fundamental}. The time-domain results are divided into two parts i.e., for single-peak potential and for double-peak potential.

\subsection{Behavior of the single-peak potential}

\begin{figure}[h]
    \includegraphics[scale=0.7]{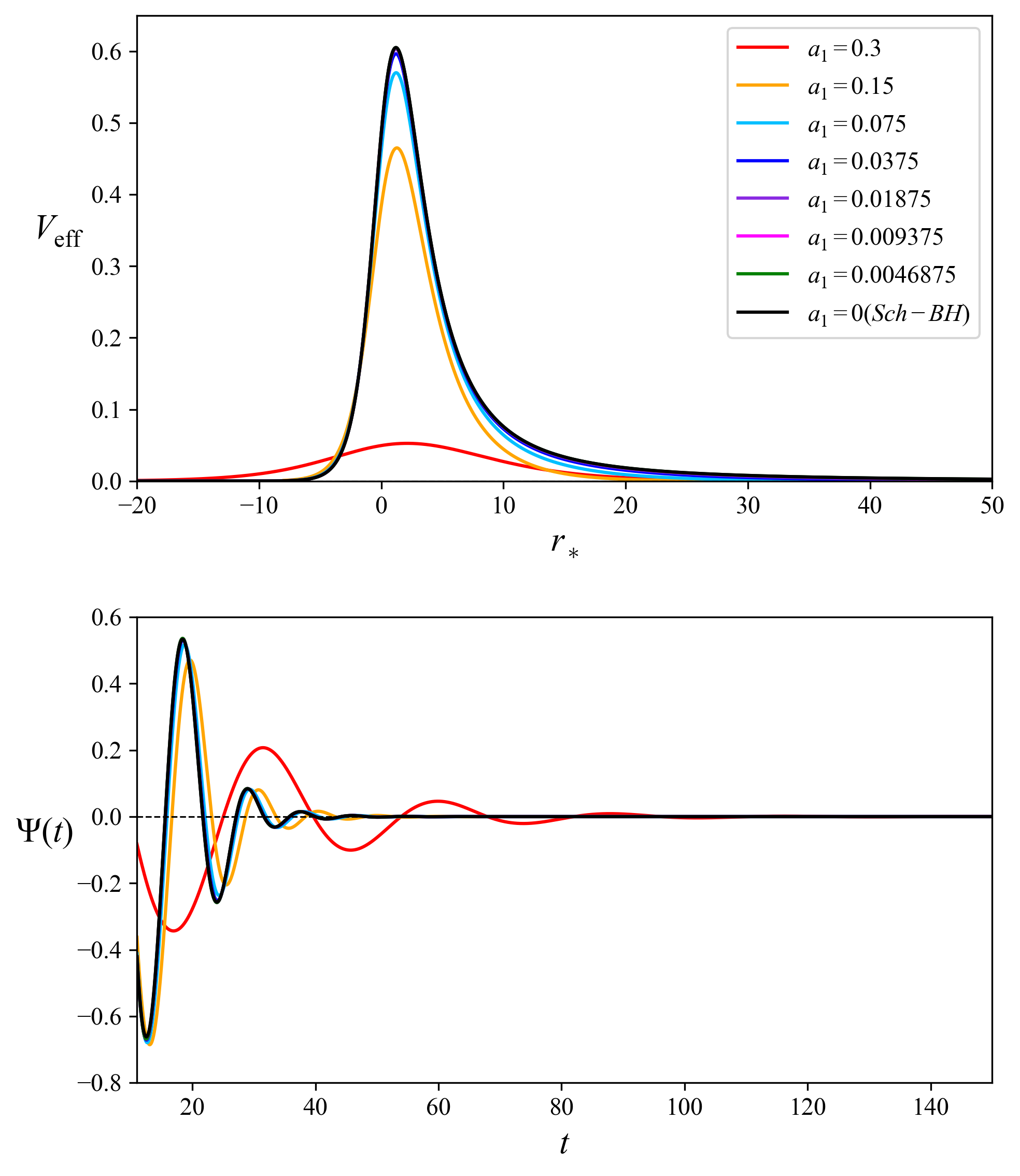}
\caption{The upper figure shows the $l=2$ effective potentials for gravitational perturbations of the three-form black hole, without the modified potential, as a function of the tortoise coordinate $r_*$. The lower figure shows the time-domain waveform for gravitational perturbation of the three-form black hole, as observed at $r_*=10$. The plots are presented for varying values of $a_1$. }
    \label{timdomain all}
\end{figure}

We present the results for the case in which the three-form fields are treated as a massless field. The effective potential from \eqref{wavefunction timedomain} includes only the gravitational perturbation potential, which can be written as $V_{\mathrm{eff}}=V_{GP}$. As shown in Fig.~\ref{timdomain all}, on the left panel, we illustrate the effective potential $V_{GP}$ as a function of tortoise coordinate for various value of $a_1$. The potential exhibits a single peak structure for several values of $a_1$. 
It is observed that at $a_1=0.0046875$, the effective potential is closely similar to the effective potential for the gravitational perturbation of a Sch-BH. 


We also determine the time-domain wave funtion resulting from the effective potential by integrating \eqref{wavefunction timedomain} using the finite difference method \eqref{gunlach} and employing \eqref{gaussian} as the initial data. As shown in the right panel of Fig.~\ref{timdomain all} , the time-domain profile at $a_1=0.0046875$ resembles  to the time-domain profile of a Sch-BH. As $a_1$ increases, the time-domain wave function demonstrates a decreasing amplitude and a slower rate of decay. We observe that the time-domain wave function at $a_1=0.3$ decays significantly slower than at other values of $a_1$.

We have evaluated the QNMs to examine how $a_1$ affects both the oscillation frequency and the rate of decay by using the 6th-order WKB and the Prony method as shown in Table~\ref{QNM all}. At small $a_1$, the real part of frequencies are closely match between the two methods. As $a_1$ increases, the difference in real part becomes more evident. We also find that, the real and imaginary part of quasinormal frequencies decrease (in magnitude for $\omega_i$) with increasing $a_1$. This means the time-domain wave function oscillates more slowly as $a_1$ increases and similar to the the decay rate. From these results, we find that in the case of a single-peak potential, the effective cosmological constant affects the oscillation of the time-domain wave function. An increasing in the effective cosmological constant results in slower oscillations and decay rate of the time-domain waveform. This is consistent with the results found in \cite{otsuki1991gravitational}.


\begin{table}
\begin{center}
    \begin{tabular}{|c|c|c|}
    \hline 
    $a_{1}$& $\text{6th-order WKB}$ &$\text{Prony}$ \\
    \hline
    0.3 & 0.198983 - 0.063607$i$ & 0.222702 - 0.058726$i$   \\
    \hline
    0.15 & 0.656713 - 0.159179$i$   &  0.655283 - 0.157871$i$  \\ 
    \hline
    0.075 & 0.725699 - 0.173452$i$ &  0.726174 - 0.170632$i$  \\
    \hline
    0.0375 & 0.741915 - 0.176717$i$ &  0.741589 - 0.174282$i$  \\
    \hline
    0.01875 & 0.745911 - 0.177517$i$ &  0.745906 - 0.174640$i$ \\
    \hline
    0.009375 & 0.746907 - 0.177716$i$ &  0.746901 - 0.174838$i$ \\
    \hline
    0.0046875 & 0.747156 - 0.177765$i$ & 0.747150 - 0.174887$i$ \\
    \hline
    0(Sch-BH) & 0.747239 - 0.177782$i$ &  0.747237 - 0.177564$i$ \\
    \hline
    \end{tabular}
\caption{The fundamental QNMs ($n=0$) for gravitational perturbations of the three-form black hole in the single-peak potential case with $M=1/2$ and $l=2$ for various values of $a_1$. The quasinormal frequencies are calculated using the 6th-order WKB method and the Prony method.} \label{QNM all}
\end{center}
\end{table}

\subsection{Behavior of the double-peak potentials}
In the previous case, we examine how $a_1$ affects the effective potential. In this case, we explore the effect of the mass of the three-form fields $(m_A)$, the coupling parameter $\lambda$ and constant $c_0$ on the effective potential, as introduced in section \ref{stuekelberg}. As shown in Fig.~\ref{0.15} on the upper panel, the effective potential in this case has a double-peak structure.
We categorize our results into three cases based on the distance between the two peaks, from close to large, each corresponding to specific values of $a_1$, namely $0.15$, $0.009375$, and $0.0046875$.

\subsubsection{$a_1=0.15$}

\begin{figure}[h]
    \centering
    \includegraphics[scale=0.7]{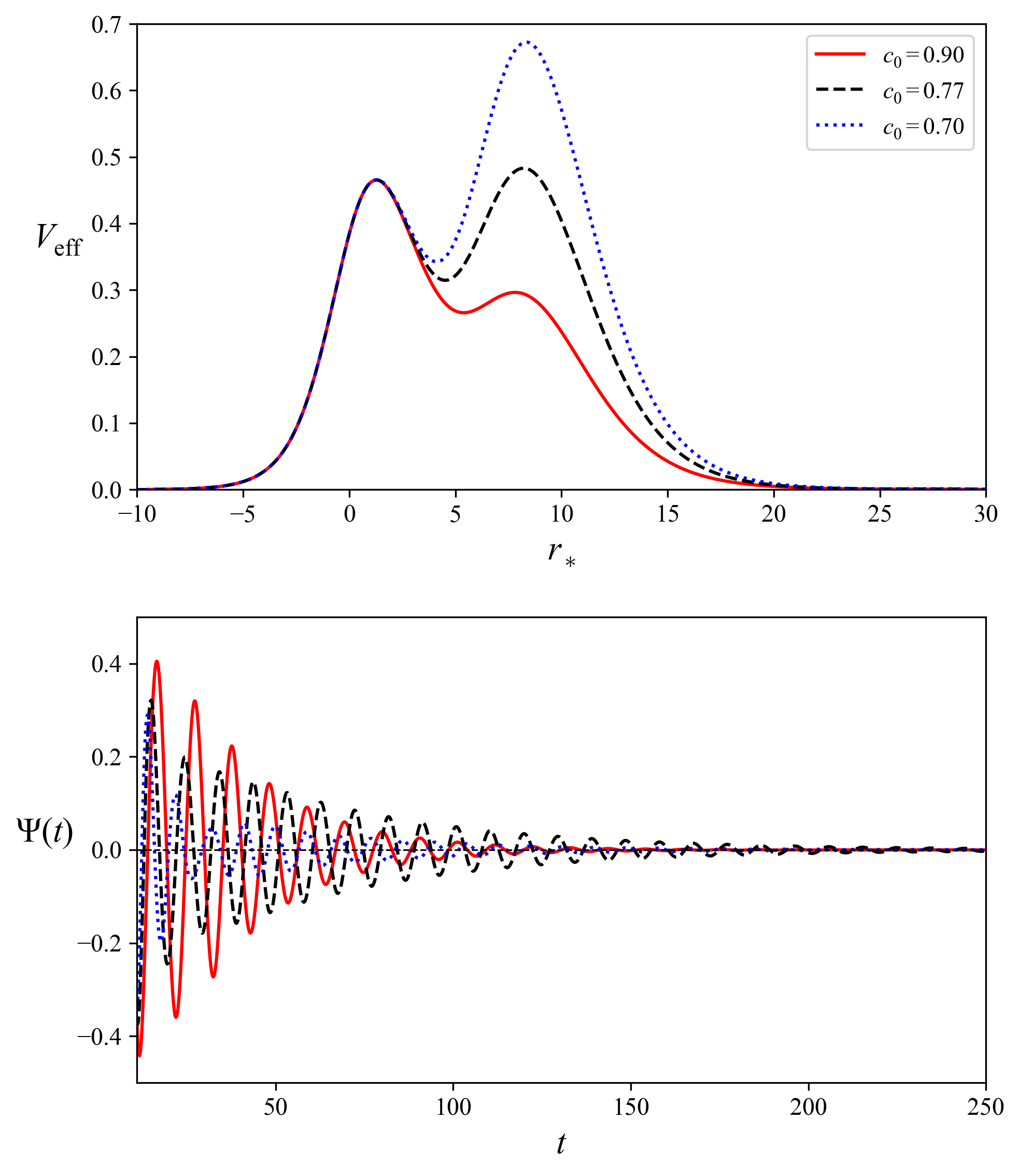}
    \caption{The (upper) $l=2$ effective potentials with modified term and the (lower) time-domain waveform for gravitational perturbations of the three-form black hole with $M=1/2$ and $a_1=0.15$, as observed at $r_*=10$. The modified potential occurs on the right hand side. In this case, we vary three values of $c_0$ : the red line represents $c_0=0.90$, the black-dashed line represents $c_0=0.77$, and the blue-dotted line represents $c_0=0.70$. The ratio of $m_A/\lambda \thicksim 10^1$} \label{0.15}
\end{figure}

In this case, the modified potential leading to the right peak while the left peak constitutes from the gravitational perturbtation potential as can be seen from the top panel of Fig.~\ref{0.15}. We remark that, in this case, we set ratio $m/\lambda \thicksim 10^1$ for all values of $c_0$. The two potentials are almost inseparable. Moreover, it is evident that the value of $c_0$ clearly influences the amplitude of both potentials. As illustrated in Fig.~\ref{0.15} on the bottom panel, the waveforms corresponding to each $c_0$ display distinct amplitudes and decay rates at late time. Their behavior is determined by the pattern of the effective potential. The waveform has the largest amplitude and fastest decay at $c_0 = 0.90$ (red line). At this value of $c_0$, the amplitude of the modified potential is larger than the amplitude of the gravitational potential. We observe that the waveform has the smallest amplitude and slowest decay at $c_0=0.70$ (blue-dotted line), where the amplitude of the interaction potential is lower than the amplitude of the gravitational perturbation potential. At $c_0=0.77$ (black-dashed line), we notice that the amplitude of the interaction potential is nearly equal to the amplitude of the gravitational perturbation potential. The waveform for this $c_0$ is represented by the black line in Fig.~\ref{0.15} on the bottom panel. We compute the corresponding frequency using the Prony method listed in Table~\ref{qnm echoes all}.

In summary, from the above results, when the black hole horizon and cosmological horizon are relatively close, it is found that 
$c_0$ is directly proportional to the amplitude of the waveform but not to the decay rate.


\subsubsection{$a_1=0.009375$}

\begin{figure}[h]
    \centering
    \includegraphics[scale=0.7]{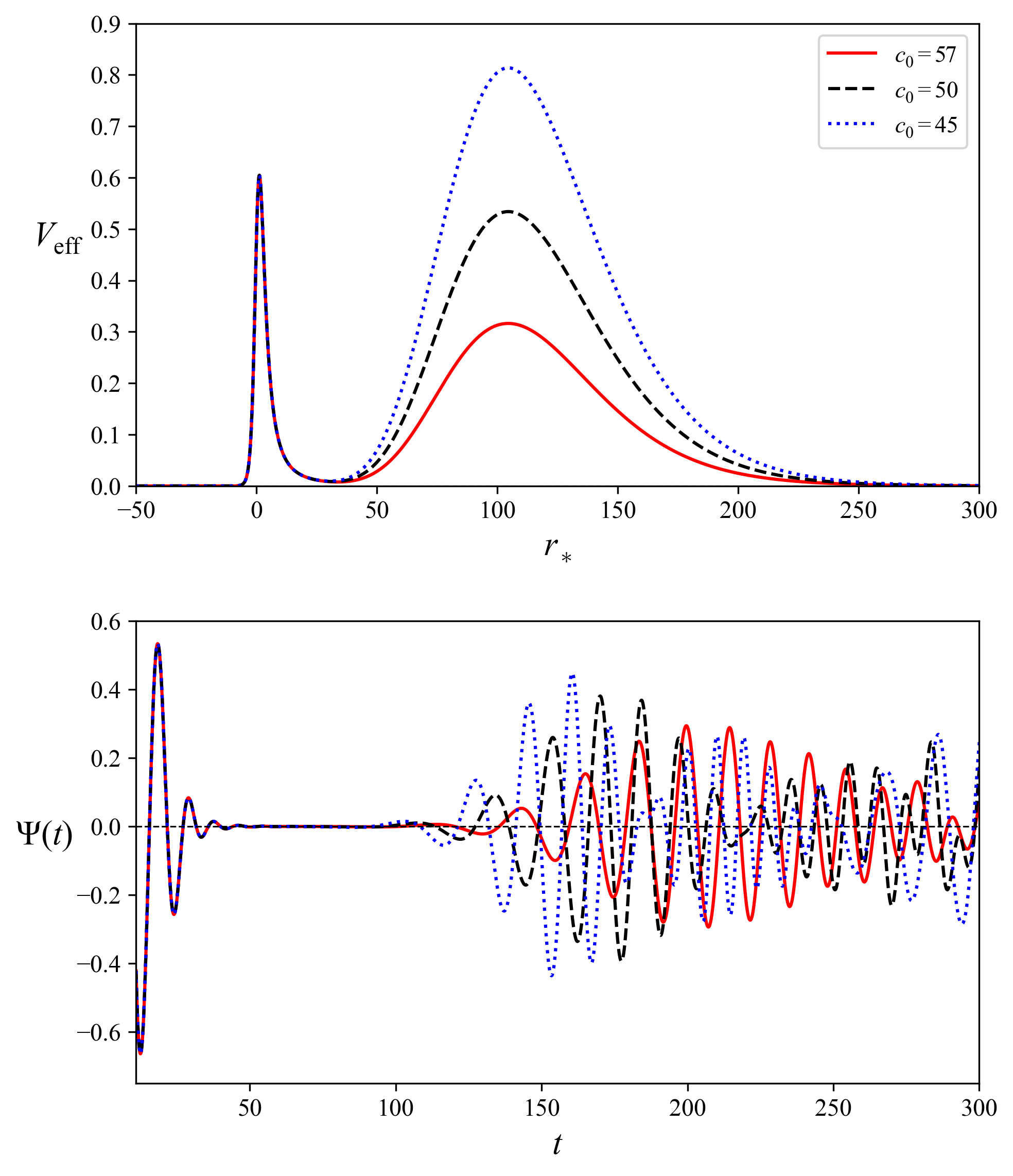}
    \caption{The (upper) $l=2$ effective potentials with interaction term and the (lower) time-domain profiles for gravitational perturbations of the three-form black hole with $M=1/2$ and $a_1=0.009375$, as observed at $r_*=10$. The modified potential occurs on the right hand side. In this case, we vary three values of $c_0$ : the red line represents $c_0=57$, the black-dashed line represents $c_0=50$, and the blue-dotted line represents $c_0=45$. The ratio of $m_A/\lambda \thicksim 10^3$.}
    \label{timdomain0009375}
\end{figure}

From the top panel of Fig.~\ref{timdomain0009375}, we observe that the modified potential appears as the right peak. Notably, the modified potential has a larger width than the gravitational perturbation potential. We vary the parameter $c_0$ to values $57, 50$ and $45$, represented by the red, black-dashed, and blue-dotted lines, respectively. Here, we fix the ratio $m/\lambda$ to $\thicksim 10^3$. As expected, only the modified potentials are affected by the mass of the three-form fields. 

In the bottom panel, we illustrate the waveform for various value of $c_0$. At early time, the waveform exhibits behavior similar to that observed in the single-peak case. At late time, however, distinctive waveforms emerge which is a consequence of the difference in value of $c_0$. For all value of $c_0$, we notice that the late time wave forms develop nearly at the same time with different amplitudes. The phase and amplitude are influenced by the amplitude of the modified potential.

In this case, we find that the amplitudes of both the modified potential and the reconstructed waveforms are proportional to $c_0$. The associated quasinormal frequencies for GWs echoes in this case are shown in Table \ref{qnm echoes all}. We observe that as $c_0$ increases, the real part of $\omega$ also increases where the decay rate decreases with $c_0$.

\subsubsection{$a_1=0.0046875$} \label{case a_1=0.0046875}
\begin{figure}[h]
    \centering
    \includegraphics[scale=0.7]{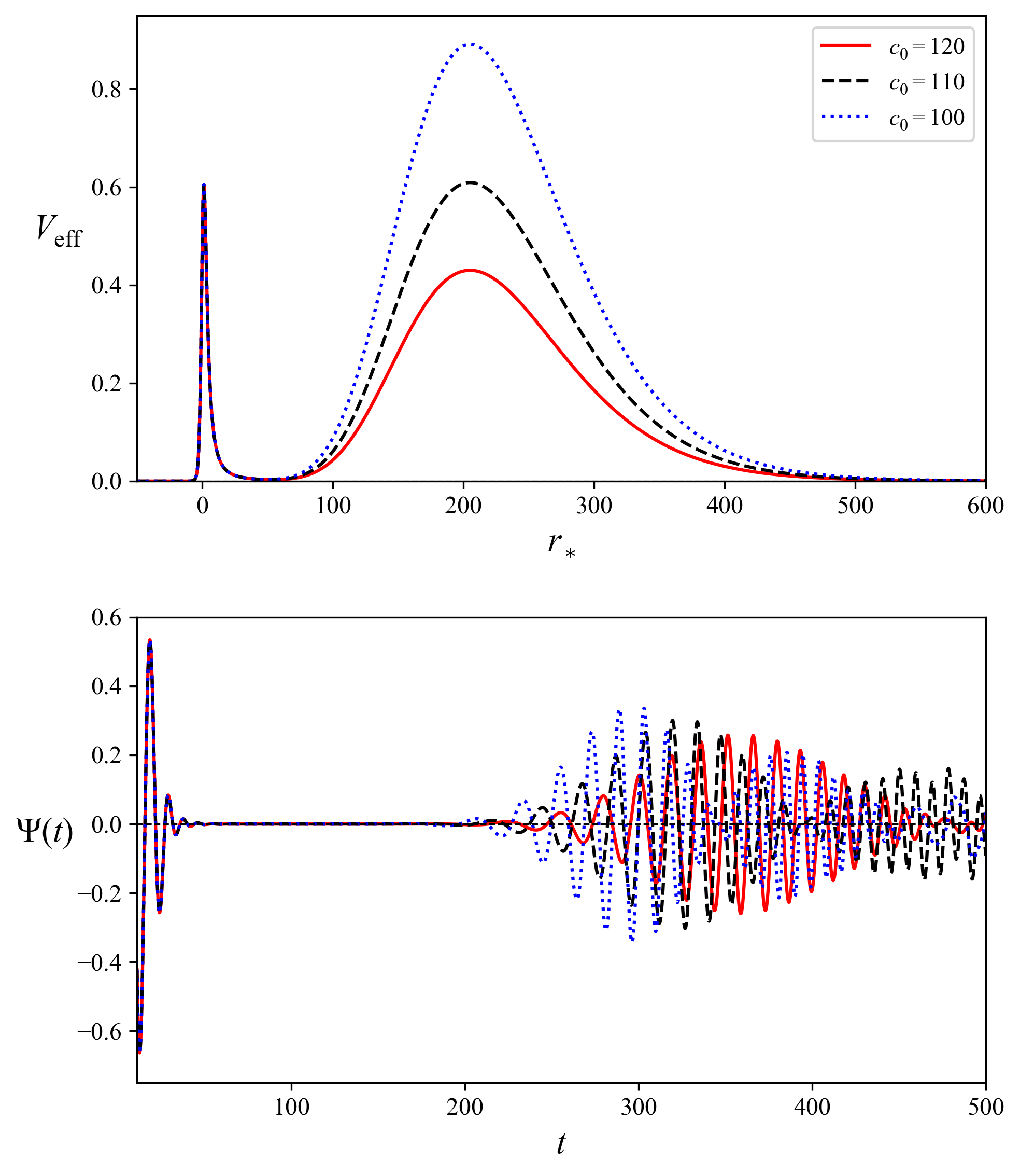}
    \caption{The (upper) $l=2$ effective potentials with interaction term and the (lower) time-domain waveform for gravitational perturbations of the three-form black hole with $M=1/2$ and $a_1=0.0046875$. The modified potential occurs on the right hand side. In this case, we vary three values of $c_0$ : the red line represents $c_0=120$, the black-dashed line represents $c_0=110$, and the blue-dotted line represents $c_0=100$. The ratio of $m_A/\lambda \thicksim 10^4$.}
    \label{timdomain 0.00468755}
\end{figure}

In this case, similar to the case of $a_1=0.009375$, the modified potential appears on the right side of the gravitational perturbation potential. Here, we have used $m_A/\lambda \thicksim 10^4$ with parameters $c_0=100,110$ and $120$. These values affect on the amplitude of the effective potential, with a noticeable impact on the modified potential. The higher the $c_0$, the smaller the amplitude of the modified potential. The time-domain profiles are shown at the bottom panel of Fig.~\ref{timdomain 0.00468755}. At early times, it is found that $c_0$ has a minimal effect on the amplitude and phase of the waveforms. At late times, $c_0$ impacts the occurrence timing of the reconstructed waveforms.

We observe that at $c_0=100,110$ and $120$, the reconstructed signal reaches its peak the at about $t=300, 330$ and $380$, respectively. In addition, second echoes signal can be seen for $c_0=100,110$ at $t=350,400$ where $c_0=120$ develops such signal at later time (not display in the plot). According to these results, we observe that the value of $c_0$ does not influence the amplitude of the reconstructed waveforms. 

In summary, we find that the time required for the reconstructed waveforms to occur is inversely proportional to the mass of the three-form fields. The associated quasinormal frequencies for the double-peak structure are calculated using the Prony method. The results for have been provided in Table \ref{qnm echoes all}.

\begin{table}
\begin{center}
\begin{tabular}{ |c|c|c|c| }
\hline
\multicolumn{4}{|c|}{$a_{1}=0.3$, $m_A/\lambda = 10^1$} \\
\hline
$c_0$& $0.70$ &$0.77$&$0.90$ \\
\hline
 $\omega$ &  0.699059 - 0.023419$i$ & 0.658883 - 0.018819$i$ & 0.393392 - 0.032323$i$ \\
\hline
\multicolumn{4}{|c|}{$a_{1}=0.0375$, $m_A/\lambda = 10^3$} \\
\hline
$c_0$& $45$ &$50$&$57$ \\
\hline
$\omega$ & 0.475379 - 0.076450$i$ & 0.481184 - 0.029184$i$ & 0.488927 - 0.008711$i$ \\
\hline
\multicolumn{4}{|c|}{$a_{1}=0.0046875$, $m_A/\lambda = 10^4$} \\
\hline
$c_0$& $100$ &$110$&$120$ \\
\hline
 $\omega$ &  0.491081 - 0.028172$i$ & 0.492917 - 0.004245$i$ & 0.493871 - 0.003375$i$ \\
\hline
\end{tabular}
\caption{The fundamental QNMs ($n=0$) for gravitational perturbations of the three-form black hole in the double-peak potential case with $M=1/2$ and $l=2$. The QNMs in this case examine the results for three values of $a_1$, with each value changing the ratio $m_A/\lambda$ and $c_0$. In this scenario, echoes occur due to the modified potential. The QNMs are calculated using the Prony method.   } \label{qnm echoes all}
\end{center}
\end{table}

\section{conclusion}
\label{conclusion}

In the present work, the gravitational perturbation of the massless three-form black hole is investigated. Static spherically symmetric black hole in Sch-dS-like form is obtained. The presence of three-form fields effectively manifests itself as a cosmological constant. The time evolution of GWs is analyzed through  gravitational perturbation by using discretization scheme. For the massless three-form fields, the gravitational perturbation potential exhibits a single-peak structure. In this case, we find that the effective cosmological constant strongly affects both the amplitude and decay rate of the time-domain waveform. In addition, we observe no echo signal. When the effective cosmological constant approaches zero, we find similar waveform as of the Sch-BH.

More importantly, the massive case of the three-form fields are also examined, with a Stueckelberg field introduced to restore gauge invariance. The equation of motion for the Stueckelberg field leads to a modification of the effective potential, resulting in a double-peak structure. The existence of a double-peak potential produces GWs echoes at late times in the time-domain profile. The echo signals are then analyzed by considering distinct values of $a_1$ and $c_0$. A smaller $a_1$ results in greater spacing between two peaks, causing the echo signal to develop more slowly than with a larger $a_1$. A higher value of parameter $c_0$ increases the potential's height, which in turn reduces the amplitude of the echo signal.


The ratio of $m_A/\lambda$ strongly affects the modified potential. To generate a double-peak effective potential, the ratio $m_A/\lambda$ must be within $10^3 - 10^4$, corresponding to $c_0$ on the order of $10^1-10^2$. The constant $c_0$ has a significant impact on the characteristics of the echoes. As a result, the behavior of the echoes depends on the distance between the black hole horizon and the cosmological horizon, as well as on the amplitudes of the modified potential. This work shows that gravitational wave echoes emerge when the black hole horizon and cosmological horizon are sufficiently distant, while no echoes are produced in the near-extremal case.
The QNMs of the three-form black hole are calculated by using the 6th-order WKB and Prony methods. We find that both methods agree very well. In addition, as $a_1$ increases the real part of the frequencies decreases while the imaginary part becomes less negative. From our analysis, we find no evidence of unstable modes. The analysis presented here is relevant to the cosmological no-hair theorem \cite{gibbons1977cosmological}. As a result, this also confirms that 
the de Sitter background is perturbative stability \cite{gibbons1983very}.

\begin{acknowledgments}
We thank Yanapat Limrachadawong for valuable discussions and contributions to exploring the methods for presenting the time-domain profiles. Special thanks also go to Nuttaphat Lunrasri for insightful discussions and for ensuring the correctness of the three-form fields analysis. NA is supported by the Research Capability Enhancement Program through a graduate student scholarship from the Faculty of Science, Khon Kaen University, which is gratefully acknowledged. SP and DS are supported by Thailand NSRF via PMU-B [grant number B39G680009]. DS has also received funding support from the Fundamental Fund of Khon Kaen University.
\end{acknowledgments}

\bibliography{ref}


\end{document}